\documentclass[12pt,a4paper]{article}
\usepackage{cite}
\usepackage{graphics}
\usepackage{color}
\usepackage{amssymb}
\usepackage{latexsym}

\begin{document}

\title{A generalised lattice Boltzmann algorithm for the flow of a nematic liquid crystal
with variable order parameter.}
\author{ C M  Care \and I Halliday  \and K Good \and S V Lishchuk\\
Materials Research Institute, Sheffield Hallam University, \\ Pond
Street, Sheffield, S1 1WB} \maketitle

\begin{abstract} A lattice Boltzmann (LB) scheme is described which recovers
the equations developed by Qian--Sheng  for the hydrodynamics of a
nematic liquid crystal with a tensor order parameter.  The
standard mesoscopic LB  scalar density is generalised to a tensor
quantity and the macroscopic momentum, density and tensor order
parameter are recovered from appropriate moments of this
mesoscopic density. A single lattice Boltzmann equation is used
with a direction dependent BGK collision term, with additional
forcing terms to recover the antisymmetric terms in the stress
tensor. A Chapman Enskog analysis is presented which demonstrates
that the Qian--Sheng scheme is recovered, provided a lattice with
sixth order isotropy is used. The method is validated against
analytical results for a number of cases including flow alignment
of the order tensor and the Miesowicz viscosities in the presence
of an aligning magnetic field. The algorithm accurately recovers
the predicted changes in the order parameter in the presence of
aligning flow, and magnetic, fields.

\end{abstract}

\section{Introduction}
\label{sec:Introduction}

The lattice Boltzmann (LB) method has been extensively studied as
a mesoscopic method of simulating isotropic fluids ({\em eg}
\cite{qian:92a, hou:95a,boghosian:98a,sacci:01a}). The strengths
of the method lie in modelling flow in complex geometries ({\em
eg} \cite{koponen:98a}) or in multi-component flow ({\em eg}
\cite{swift:96a,halliday:98a}). Recently, a number of LB schemes
have been developed to represent the flow of anisotropic fluids
such as liquid crystals \cite{care:00a,denniston:01a}.

Materials which exhibit liquid crystal phases have anisometric
molecules \cite{deGennes:93a,allen:93a} and nematic liquid
crystals are of particular interest because of their application
in display devices. There is also increasing interest in modelling
liquid crystal colloids in which colloidal particles are embedded
in a nematic phase \cite{stark:01a}.  The colloidal particles
interact through the distortions and defects which they generate
in the nematic elastic field. The forces between the particles and
the dynamics of their motion is accessible to experiment ({\it cf}
Poulin {\em et al} \cite{poulin:97a}).  As a consequence of these
additional colloidal interactions the particles rearrange
themselves into new structures and the LB method provides a
particularly effective way of dealing with the complex boundary
conditions in such problems. The method can recover both the
nematostatics and nematodynamics of these phases and hence the
dynamics of the colloidal phase can be captured and this has been
achieved with an earlier LB method developed by the authors
\cite{care:00a,care:02-C46}. Additionally, the colloidal particles
may be an isotropic fluid and work is currently in progress to
extend the two phase LB algorithms which have been developed for
isotropic fluids to describe a mixture in which one of the phases
is a nematic liquid crystal. This current work is based on the
algorithms presented in this paper extended to embody the
isotropic-nematic interface following the macroscopic description
of Rey \cite{rey:00a,rey:01a}. The hydrodynamics associated with
pair annihilation of line defects is also a matter of current
interest and results have been reported using both lattice
Boltzmann \cite{toth:02-105504} and conventional solvers
\cite{svensek:02-021712} of the equations for nematodynamics in
the presence of a variable order parameters.

The orientational ordering of the molecules in the nematic phase
is characterised by a director field, $n_{\alpha}({\bf x}, t)$, a
unit vector which essentially defines the \lq average orientation'
of the molecules. However, the nematic ordering is more fully
characterised by a traceless and symmetric order tensor,
$Q_{\alpha\beta}$. In this work, for simplicity, we assume that
the director is confined to a two dimensional plane, and hence
that $Q_{\alpha\beta}$  may be written  in the form

\begin{equation}\label{eq:3DOrderTensor}
  Q_{\alpha\beta}({\bf x}, t)=S (2 n_{\alpha} n_{\beta}-\delta_{\alpha\beta})
\end{equation}
The principal eigenvector of $Q_{\alpha\beta}$ is the director and
the principal eigenvalue, $S({\bf x}, t)$, is the scalar order
parameter.

The scheme proposed by Care {\em et al} \cite{care:00a} to recover
the Ericksen, Leslie and Parodi (ELP) equations
    \cite{deGennes:93a} used two coupled LB equations.  These
equations govern the continuum fluid dynamics of an incompressible
nematic with an order parameter, $S$, which is both position and
time \textit{independent}. In the scheme \cite{care:00a}, one of
the LB equations carried the momentum and the second carried a
vector density corresponding to the director field. The Denniston
{\em et al} \cite{denniston:01a} scheme is also based on two
coupled lattice Boltzmann schemes one of which carries a momentum
density and the second of which carries an tensor density from
which the macroscopic order tensor can be recovered. This latter
scheme recovers the Beris--Edwards equations \cite{beris:94a} for
the flow of a nematic liquid crystal with variable order
parameter.

In the work presented in this paper, a third scheme is presented
which recovers the Qian--Sheng \cite{qian:98a} equations for the
flow of a nematic liquid crystal with a variable scalar order
parameter. The scheme is based on a single LB equation which
governs the evolution of a tensor density and from which both the
macroscopic order and momentum evolution equations are recovered.
It is assumed that the dynamics generated by the Qian--Sheng
equations are essentially equivalent to those generated by the
Beris-Edwards equations \cite{beris:94a} although this equivalence
has not been demonstrated explicitly. Recent work by Sonnet,
Maffettone and Virga \cite{sonnet:02a} provides the basis upon
which the variety of schemes with a variable order parameter may
be compared. The Qian--Sheng equations, like the Beris--Edwards
equations, reduces to the ELP formalism in the limit that the
order parameter becomes independent of time and position. There
are a number of important differences between the scheme proposed
here and that of Denniston. In the scheme described here, the
target equations are those of Qian--Sheng rather than
Beris--Edwards, there is a single Boltzmann equation, the
equilibrium distribution function is isotropic (as is expected on
physical grounds), and the scheme is able to recover the full
tensorial coupling of the order tensor to the velocity gradient
tensor, from a single lattice distribution function.

The target macroscopic equations of the Qian--Sheng scheme are
summarised in Section \ref{sec:MacroscopicScheme},  the new scheme
is described in Section \ref{sec:TheNewScheme} and a Chapman
Enskog analysis of the scheme is presented in Section
\ref{sec:AnalyisOfAlgorithm}.  A number of analytical results are
developed from the Qian--Sheng equations in Section
\ref{sec:AnalyticalResults}, which provide the basis for
validation of the LB algorithm in Section \ref{sec:Results}.  The
conclusions are presented in Section \ref{sec:Conclusions}.

\section{Qian-Sheng formalism}
\label{sec:MacroscopicScheme} In this section we summarise the
target macroscopic equations for the LB method.  The two governing
equations of the Qian scheme are the momentum evolution equation
\begin{equation}
\label{eq:qianMomentum}
 \rho D_{t}u_{\beta} =
  \partial_{\beta} (- P \delta_{\alpha \beta}+ \sigma^{d}_{\alpha \beta } + \sigma^f_{
     \alpha \beta} + \sigma'_{\alpha \beta} )
\end{equation}
and the order tensor evolution equation
\begin{equation}
\label{eq:qianOrder}
 J \ddot{Q}_{\alpha\beta}=h_{\alpha\beta}+h'_{\alpha\beta}-\lambda\delta_{\alpha\beta}
-\varepsilon_{\alpha\beta\gamma}\lambda_{\gamma}
\end{equation}
It is shown by Qian that in the limit of constant order parameter,
the solutions of these equations is identical to those obtained
from the ELP equations. Throughout this work we use the repeated
index notation for summations over Cartesian indices. In the above
equations $D_{t}=\partial_{t}+u_{\mu}\partial_{\mu}$ is the
convective derivative, $P$ is the pressure, $\sigma^{d}_{ \alpha
\beta}$ is the distortion stress tensor given by
\begin{equation}\label{eq:distortionStressTensor}
\sigma^{d}_{ \alpha \beta}=- \frac{\partial
F_{LdG}}{\partial\,(\partial_{\alpha Q_{\mu\nu}})}
\partial_{\beta}Q_{\mu\nu}
\end{equation}
and $\sigma^f_{\alpha \beta}$ is the stress tensor associated with
an externally applied field.  In the current work we only consider
an externally applied magnetic field for which the stress tensor
is given by Landau and Lifshitz \cite{landau:84a}
\begin{equation}\label{eq:fieldStressTensor}
\sigma^f_{\alpha \beta}=-\frac{1}{4 \pi}(H_{\alpha}
H_{\beta}-\frac{1}{2}H^{2}\delta_{\alpha\beta})
\end{equation}
The viscous stress tensor $\sigma'_{\alpha \beta}$ is given by
\begin{eqnarray}\label{eq:viscousStressTensor}
\sigma'_{\alpha \beta} &= &
\beta_{1}Q_{\alpha\beta}\,Q_{\mu\nu}A_{\mu\nu}+\beta_{4}A_{\alpha\beta}
  + \beta_{5}Q_{\alpha\mu}A_{\mu\beta}+
\beta_{6}Q_{\beta\mu}A_{\mu\alpha} + \nonumber \\
  & &
\frac{1}{2}\mu_{2}N_{\alpha\beta}-\mu_{1}Q_{\alpha\mu}N_{\mu\beta}
+\mu_{1}Q_{\beta\mu}N_{\mu\alpha}
\end{eqnarray}
where
$A_{\alpha\beta}=\frac{1}{2}(\partial_{\alpha}u_{\beta}+\partial_{\beta}u_{\alpha})$
is the symmetric velocity gradient tensor. The elastic molecular
field is given by
\begin{equation}\label{eq:elasticMolecularField}
  h_{\alpha\beta}=-\frac{\partial F_{LdG}}{\partial\, Q_{\alpha\beta}}
  +\partial_{\mu}\frac{\partial F_{LdG}}{\partial\,(\partial_{\mu} Q_{\alpha\beta})}
\end{equation}
where, in the current work, the Landau-deGennes free energy is
assumed to be of the form
\begin{eqnarray}\label{eq:LandauFreeEnergy}
  F_{LdG}& =& \frac{1}{2}(\alpha_{F} Q_{\mu\nu}^{2} + L_{1}Q_{\mu\nu,\tau}^{2}+
  L_{2}Q_{\mu \nu,\nu}Q_{\mu \tau,\tau}
   ) - \nonumber \\ & &
   \beta_{F} Q_{\mu\nu}Q_{\nu\tau}Q_{\tau\mu}+\gamma_{F} (Q_{\mu\nu}^{2})^{2}
\end{eqnarray}
where $Q_{\alpha \beta,\gamma}\equiv\partial_{\gamma}(Q_{\alpha
\beta})$. Using this form of the free energy and the definition of
the molecular field, equation (\ref{eq:elasticMolecularField}), we
find
\begin{equation}\label{eq:MolecularField}
  h_{\alpha\beta}=L_{1}\partial_{\mu}^{2}Q_{\alpha\beta}+L_{2}\partial_{\beta}
  \partial_{\mu}Q_{\alpha\mu}-\alpha_{F}Q_{\alpha\beta}+ 3 \beta_{F}Q_{\alpha\mu}Q_{\beta\mu}
  - 4 \gamma_{F}Q_{\alpha\beta} Q_{\mu\nu}^{2}
\end{equation}
The quantities $\lambda$ and $\lambda_{\alpha}$ in equation
(\ref{eq:qianOrder}) are Lagrange multipliers which impose the
constraints on the elastic molecular field which arise because the
order tensor, $Q_{\alpha\beta}$, is symmetric and traceless. For a
three dimensional system they have the values
\begin{equation}
\begin{array}{ccccc}\label{eq:LagrangeMultipliers}
\lambda = \frac{1}{3}(h_{\mu\mu}-\frac{1}{2}\mu_{2}A_{\mu\mu}) &
~~~~~~ &
 \lambda_{\alpha} & = &
\frac{1}{2}\varepsilon_{\alpha\mu\nu}h_{\mu\nu}
\end{array}
\end{equation}
The term $A_{\mu\mu}$ in the first of equations
(\ref{eq:LagrangeMultipliers}) does not appear in the Qian-Sheng
equations, but is necessary in order to correct for the slight
compressibility of the LB fluid.  In the presence of an external
magnetic field the free energy is assumed to be augmented by a
term of the form
\begin{equation}\label{eq:MagneticFreeEnergy}
  F_{H}=-\frac{1}{2}(\chi_{\parallel}-\chi_{\perp})Q_{\alpha\beta}H_{\alpha}H_{\beta}
\end{equation}
which gives an additional  term in the molecular field of the form
\begin{equation}\label{eq:MagenticMolecularField}
  h_{\alpha\beta}^{H}=\frac{1}{2}\chi_{a}H_{\alpha}H_{\beta}
\end{equation}
where $\chi_{a}=\chi_{\parallel}-\chi_{\perp}$ is the  anisotropy
in the susceptibility. The viscous molecular field,
$h'_{\alpha\beta}$, is given by
\begin{equation}\label{eq:viscousMolecularField}
  h'_{\alpha\beta}=\frac{1}{2}\,\mu_{2}A_{\alpha\beta}+\mu_{1}N_{\alpha\beta}
\end{equation}
where $N_{\alpha\beta}$ is the co-rotational derivative defined by
\begin{equation}\label{eq:Ndefinition}
N_{\alpha\beta}=\partial_{t}Q_{\alpha\beta}+u_{\mu}
\partial_{\mu}Q_{\alpha\beta}-\varepsilon_{\alpha\mu\nu}\omega_{\mu}Q_{\nu\beta}-
\varepsilon_{\beta\mu\nu}\omega_{\mu}Q_{\nu\alpha}
\end{equation}
where the vorticity, $\underline{\omega}=\frac{1}{2}\,(\nabla
\times \underline{u})$.

In order to develop a lattice Boltzmann scheme we re-arrange
equation (\ref{eq:viscousStressTensor}) by substituting from
equation (\ref{eq:Ndefinition}) for the co-rotational derivative,
$N_{\alpha\beta}$, and obtain the form
\begin{eqnarray}\label{eq:ViscousStressTensorRearranged}
\sigma'_{\alpha \beta} &= &
\beta_{1}Q_{\alpha\beta}\,Q_{\mu\nu}A_{\mu\nu}+(\beta_{4}-\frac{(\mu_{2})^{2}}{2
\mu_{1} })A_{\alpha\beta}
  + (\beta_{5}+\frac{1}{2}\mu_{2})Q_{\alpha\mu}A_{\mu\beta}+\nonumber \\
  & & (\beta_{6}-\frac{1}{2}\mu_{2})Q_{\beta\mu}A_{\mu\alpha} + \frac{\mu_{2}}{2
  \mu_{1}}(h_{\alpha\beta}-\delta_{\alpha\beta} \lambda-
  \varepsilon_{\alpha\beta\mu}\lambda_{\mu})+\nonumber\\
  & &  Q_{\beta\mu}h_{\mu\alpha}-
  Q_{\alpha\mu}h_{\mu\beta} \varepsilon_{\mu\beta\nu}Q_{\alpha\mu}\lambda_{\nu}-
  \varepsilon_{\mu\alpha\nu}Q_{\beta\mu}\lambda_{\nu}
\end{eqnarray}
Equation (\ref{eq:qianOrder}) for the order tensor evolution can
be cast in the form
\begin{eqnarray}\label{eq:qianOrderRearranged}
 D_{t}Q_{\alpha\beta}& =&  -\frac{\mu_{2}}{2 \mu_{1}}A_{\alpha\beta}+
 \frac{1}{\mu_{1}} h_{\alpha\beta} + \epsilon_{\alpha\mu\nu}\omega_{\mu}Q_{\nu\beta}+
 \epsilon_{\beta\mu\nu}\omega_{\mu}Q_{\nu\alpha}\nonumber \\
 & & -\frac{1}{\mu_{1}}(\delta_{\alpha\beta}\lambda+
 \varepsilon_{\alpha\beta\mu}\lambda_{\mu})
\end{eqnarray}
where we have set the inertial density of the fluid, $J$, to zero.
Equations (\ref{eq:ViscousStressTensorRearranged}) and
(\ref{eq:qianOrderRearranged}) are the expressions which are
recovered by the LB scheme proposed below.  After each time step
the macroscopic density, momentum and the order tensor are
recovered from the LB tensor density. The gradients of the order
tensor are then used to construct the molecular field through
equation (\ref{eq:MolecularField}) and this modifies the dynamics
through appropriate forcing terms in the LB equation.

\section{The lattice Boltzmann algorithm}
\label{sec:TheNewScheme} In order to recover both the momentum
evolution equation (\ref{eq:qianMomentum}) and the order tensor
evolution equation (\ref{eq:qianOrder}) within a lattice Boltzmann
scheme, the scalar density of a standard lattice Boltzmann scheme,
$f_{i}(\underline{r},t)$, is replaced by a tensor density
$g_{i\alpha\beta}(\underline{r},t)$ where $i$ is the normal
velocity index and $\alpha$ and $\beta$ label either a two, or
three, dimensional Cartesian basis. One can think of the tensor
density as carrying information about the ordering of that
population of the fluid \lq element' on the velocity link $i$,
associated with a particular position and time. Hence the
densities of a standard LB scheme have been generalised to carry
information about the order associated with the fluid in addition
to the density and momentum.

The density $g_{i\alpha\beta}(\underline{r},t)$ is assumed to have
the following moments
\begin{eqnarray}\label{eq:densityMoment}
  \rho & = & \sum_{i}g_{i\mu\mu}\\
  \rho u_{\alpha}& = &\sum_{i}c_{i\alpha}g_{i\mu\mu}\\
\rho S_{\alpha\beta}& = & \sum_{i}g_{i\alpha\beta}
\end{eqnarray}
where $S_{\alpha\beta}$ is an order matrix with unit trace which
is related to $Q_{\alpha\beta}$ by
\begin{equation}\label{eq:SOrderTensor}
S_{\alpha\beta}=\frac{1}{2}(Q_{\alpha\beta}+\delta_{\alpha\beta})
\end{equation}
where $d$ is the dimensionality of the space which the Cartesian
indices $\alpha$ and $\beta$ span. The LB algorithm governing the
evolution of $g_{i\alpha\beta}$ is taken to be of the form
\begin{equation}\label{eq:BoltzmannEquation}
g_{i\alpha\beta}(\underline{r} + \delta \underline{c}\,_{i},t +
\delta )=g_{i\alpha\beta}(\underline{r},t)+
\sum_{j}g_{j\mu\nu}^{(neq)}(\underline{r},t) M_{i\alpha\beta j
\mu\nu} + \phi_{i\alpha\beta} +\chi_{i\alpha\beta}
\end{equation}
where $\phi_{i\alpha\beta}$ and $\chi_{i\alpha\beta}$ are the
forcing terms for the momentum and order respectively and there is
an implicit summation over repeated Cartesian indices. The
non-equilibrium distribution function is given by
\begin{equation}\label{eq:NonEquilibriumDistributionFunction}
g_{i\alpha\beta}^{(neq)}=g_{i\alpha\beta}-g_{i\alpha\beta}^{(0)}
\end{equation}
where the equilibrium distribution function,
$g_{i\alpha\beta}^{(0)} $, is taken to be of the form
\begin{equation}\label{eq:EquilibriumDistributionFunction}
  g_{i\alpha\beta}^{(0)}=S_{\alpha\beta}f_{i}^{(0)}.
\end{equation}
The distribution function, $f_{i}^{(0)}$, is assumed to be second
order in the velocity $u_{\alpha}$ and is determined in the usual
way, by the requirements
\begin{equation}\label{eq:EquilibriumDistributionConditions}
\rho  =  \sum_{i} f_{i}^{(0)} ~~~~~~~~~~~~~ \rho u_\alpha   =
\sum_{i} c_{i \alpha} f_{i}^{(0)} \label{eq:moments}
\end{equation}
and Galilean invariance in the form
\begin{equation}\label{eq:GalileanInvariance}
  \sum_{i}c_{i\alpha}c_{i\beta}g_{i\alpha\beta}^{(0)}= c_{s}^{2}\rho \delta_{\alpha\beta}
  +\rho u_{\alpha} u_{\beta}
\end{equation}
Hence, we have
\begin{equation}\label{eq:EquilibriumDistributionFunctionFirstMoment}
\sum_{i}g_{i\alpha\beta}^{(0)}=\rho S_{\alpha\beta}~~~~~~~~~~~
\sum_{i}c_{i\alpha}g_{i\alpha\beta}^{(0)}=\rho
S_{\alpha\beta}u_{\alpha }
\end{equation}

In Section \ref{sec:AnalyisOfAlgorithm}, it is shown that in order
to recover the required tensor coupling of the order tensor to the
velocity gradient tensor it is necessary to have sixth order
isotropy of the velocity tensors. In this work, the required
isotropy is achieved by using a $D2Q13$ three speed hexagonal
lattice with velocity vectors, ${\bf c}_p$, given by
\begin{eqnarray}
{\bf c_0} & = &  \{ 0,0 \} \nonumber \\ {\bf c_1} & = & c \{ \pm
1, 0 \}, c \{ \pm 1/2, \pm \surd 3 /2 \} \nonumber \\ {\bf c_2} &
= & c \{ 0, \pm \surd 3 \}, c\{ \pm 3/2, \pm \surd 3 /2 \}
\end{eqnarray}
the subscripts 0, 1 and 2 being associated with particles with
velocity 0, $c$ and $\surd 3 c$ respectively. Imposing the
conditions (\ref{eq:EquilibriumDistributionConditions}) and
(\ref{eq:GalileanInvariance}) leads to the result
\begin{equation}\label{eq:EquilibriumDistributionFunctionf}
g_{i\alpha\beta}^{(0)}=  \rho S_{\alpha\beta} t_{i} \left[  1 +
\frac{1}{c_s^2} u_\alpha c_{i \alpha} + \frac{1}{2 c_s^2} u_\alpha
u_\beta \left( \frac{c_{i \alpha} c_{i \beta}}{c_s^2} -
\delta_{\alpha \beta} \right) \right]
\end{equation}
where, in $D2Q13$, $ t_0  =   11/25,  t_1 = 9/100$ and $t_2 = 1 /
300$ and the velocity of sound $c_s = (3/10)^{1/2}$. The collision
operator is taken to be of the form
\begin{equation}\label{eq:collisionOperator}
M_{i\alpha\beta j \mu\nu} =\delta_{\alpha
\mu}\delta_{\beta\nu}\left( - \frac{\delta_{i,j}}{\tau_{j}}+
\frac{\delta_{i 0 }}{\tau_{j}}+ \frac{t_{i} c_{i\varepsilon}
c_{j\varepsilon} }{c_{s}^{2}}(\frac{1}{\tau_{j}}-2)\right)
\end{equation}
where the {\em direction dependent} relaxation parameter is
defined to be
\begin{equation}\label{eq:relaxtionParameter}
\tau_{i}=\tau_{0}(1+\eta_{0}c_{i\mu}c_{i\nu}\delta_{\mu\nu}+
\eta_{2}c_{i\mu}c_{i\nu}Q_{\mu\nu})
\end{equation}
This collision operator may be seen more transparently after
performing the contraction with the non-equilibrium distribution
function as in equation (\ref{eq:BoltzmannEquation})
\begin{equation}\label{eq:collisionOperatorLBGKForm}
\sum_{j\mu\nu}M_{i\alpha\beta j
\mu\nu}g_{j\mu\nu}^{(neq)}=-\frac{g_{i\alpha\beta}^{(neq)}}{\tau_{i}}+
\delta m_{i\alpha \beta} +\delta p_{i\alpha\beta}
\end{equation}
where
\begin{eqnarray}\label{eq:LBGKcorrectionsMass}
\delta m_{i\alpha \beta}& = &\delta_{i
0}\sum_{j}\frac{g_{j\alpha\beta}^{(neq)}}{\tau_{j}}\\
\end{eqnarray}
is a correction to conserve mass.  The term
\begin{eqnarray}\label{eq:LBGKcorrectionsMomentum}
 \delta
p_{i\alpha\beta} & = & \frac{t_{i}
c_{i\varepsilon}}{c_{s}^{2}}\sum_{j}c_{j\varepsilon}g_{j\alpha\beta}^{(neq)}\left
(\frac{1}{\tau_{j}}-2\right)
\end{eqnarray}
includes a term in $1/\tau_{j}$ which is a correction to conserve
momentum and a term associated with a factor $-~2$ which is
explained at the end of Section \ref{sec:ChapmanEnskog}. The
collision operator is essentially an LBGK collision operator ({\em
eg} \cite{hou:95a}) with a relaxation time which is direction
dependent. When written in the matrix form, the collision operator
is seen to lie between a conventional LBGK operator and a
linearised lattice Boltzmann scheme ({\em eg} \cite{higuera:89a})
although the usual circulant properties of the matrix associated
with an isotropic fluid are destroyed by the direction dependent
scattering in the method presented here.

The momentum forcing term is given by
\begin{equation}\label{eq:momentumForcing}
\phi_{i\alpha\beta}=t_{i} S_{\alpha\beta}\, c_{i\mu}
\partial_{\nu}F_{\nu\mu}
\end{equation}
where
\begin{eqnarray}\label{eq:FAlphaBeta}
  \lefteqn{ F_{\alpha\beta}=-\frac{1}{c_{s}^{2}}\left( -
  \frac{\mu_{2}( h_{\alpha\beta}-\varepsilon_{\alpha\beta\mu}\lambda_{\mu})}{2 \mu_{1
  }}\right.}\nonumber\\ & & \left. +
  Q_{\alpha\mu}(h_{\mu\beta}-\varepsilon_{\mu\beta\nu}\lambda_{\nu})-
  Q_{\beta\mu}(h_{\mu\alpha}-\varepsilon_{\mu\alpha\nu}\lambda_{\nu})-
  \sigma_{\alpha\beta}^{d}-\sigma_{\alpha\beta}^{f}\right)
\end{eqnarray}
and the angular forcing term is given by
\begin{eqnarray}\label{eq:angularForcing}
\lefteqn{\chi_{i\alpha\beta}=t_{i}\rho
\left[(\varepsilon_{\alpha\nu\mu}Q_{\beta\mu}+
\varepsilon_{\beta\nu\mu}Q_{\alpha\mu})\omega_{\nu}\right]
 -}\nonumber\\& &\frac{t_{i}\rho}{2
\mu_{1}}\left[\mu_{2}A_{\alpha\beta}-2 h_{\alpha\beta} + 2
\delta_{\alpha\beta}\lambda+2
\varepsilon_{\alpha\beta\mu}\lambda_{\mu} \right]
\end{eqnarray}
\section{Analysis of the algorithm}
\label{sec:AnalyisOfAlgorithm}
\subsection{Chapman Enskog expansion} \label{sec:ChapmanEnskog} In this section the key
results of the Chapman Enskog analysis are given which demonstrate
how the scheme described in Section \ref{sec:MacroscopicScheme}
can be recovered from the LB algorithm described in Section
\ref{sec:TheNewScheme}.  We follow a standard Chapman Enskog
analysis (\cite{hou:95a}) and expand the density and the time
derivatives in the form
\begin{eqnarray}\label{eq:CEDensityExpansion}
  g_{i\alpha\beta}& =& \sum_{n=0}^{\infty}\delta
  ^{n}g_{i\alpha\beta}^{(n)}\\
  \partial_{t}& = &\sum_{n=0}^{\infty} \delta^{n} \partial_{t,n}
\end{eqnarray}
We assume for the purposes of the following analysis that the
forcing terms, $\phi_{i\alpha\beta}$ and $\chi_{i\alpha\beta}$,
can both be introduced at order $O(\delta^{2})$ since both include
gradient terms in both the velocity and director field ({\em
cf}\cite{care:00a}).  The choice is supported by the agreement of
the Chapman Enskog analysis with the measured simulation data;
however the assumption relies upon assumptions implicit in LB
hydrodynamics in general.  Accordingly, it requires more careful
analysis which will be undertaken in later work.

The Chapman Enskog expansion gives to $O(\delta)$
\begin{equation}\label{eq:ChapmanEnskog1}
  \partial_{t,0}g_{i\alpha\beta}^{(0)}+c_{i\mu}\partial_{\mu}g_{i\alpha\beta}^{(0)}=
  \sum_{j}g_{j\mu\nu}^{(1)}M_{i\alpha\beta j\mu\nu}
\end{equation}
and to $O(\delta^{2})$
\begin{eqnarray}\label{eq:ChapmanEnskog2}
\protect\lefteqn{
\partial_{t,1}g_{i\alpha\beta}^{(0)}+(\partial_{t,0}+c_{i\mu}\partial_{\mu})\left\{
g_{i\alpha\beta}^{(1)}+\frac{1}{2}\sum_{j}g_{j\mu\nu}^{(1)}M_{i\alpha\beta
j\mu\nu}\right\}  = } \hspace{2.0in} \nonumber \\
 & &  \sum_{j}g_{j\mu\nu}^{(2)}M_{ i \alpha \beta j \mu\nu}
+\phi_{i\alpha\beta} +\chi_{i\alpha\beta}
\end{eqnarray}
We now take moments of equations (\ref{eq:ChapmanEnskog1}) and
(\ref{eq:ChapmanEnskog2}) in order to recover the macroscopic
equations to which the algorithm is equivalent. If we sum the two
equations over the index $i$, take the trace of the distribution
density and sum the resulting equations, we recover
\begin{equation}\label{eq:CEZeroMoment}
(\partial_{t,0} +\partial_{t,1}) \rho+ \partial_{\mu}(\rho
u_{\mu}) =0
\end{equation}
which is the continuity equation to second order.  If we multiply
equations (\ref{eq:ChapmanEnskog1}) and
(\ref{eq:CEDensityExpansion}) by $ c_{i\nu}$, sum over the index
$i$, take the trace of the distribution density and sum the
resulting equations we recover
\begin{equation}\label{eq:CEFirstMoment}
 (\partial_{t,0}+\partial_{t,1})(\rho u_{\alpha})+\partial_{\mu}
 (\Pi_{\alpha\mu}^{(0)}+\Pi_{\alpha\mu}^{(1)}+\frac{1}{2}\Omega_{\alpha\mu}^{(1)}) =
 c_{s}^{2}\partial_{\mu}F_{\mu\alpha}
\end{equation}
where
\begin{equation}\label{eq:PiDefinition}
  \Pi_{\alpha\beta}^{(n)}=\sum_{i}c_{i\alpha}c_{i\beta}g_{i\mu\mu}^{(n)}
\end{equation}
and
\begin{equation}\label{eq:OmegaDefinition}
  \Omega_{\alpha\beta}^{(n)}=\sum_{i j}g_{j\mu\nu}^{(n)}c_{i\alpha}c_{i\beta}
  M_{i\kappa\kappa j \mu\nu}
\end{equation}
Further progress can be made by noting that, to $O(u)$, the lowest
order equation (\ref{eq:ChapmanEnskog1}) can be written in the
form
\begin{equation}\label{
eq:CE1DerivativeForm} t_{i}\left( \frac{c_{i \mu}c_{i
\nu}}{c_{s}^{2}}-\delta_{\mu\nu} \right)\partial_{\mu}(\rho
S_{\alpha\beta} u_{\nu})=\sum_{j}g_{j\mu\nu}^{(1)}M_{i \alpha\beta
j \mu\nu}
\end{equation}
Substituting into equation (\ref{eq:OmegaDefinition}), this yields
the result
\begin{equation}\label{eq:OmegaDefinition2}
  \Omega_{\alpha\beta}^{(1)}=\sum_{i}t_{i}c_{i\alpha} c_{i \beta}\left(
  \frac{c_{i\mu}c_{i \nu}}{c_{s}^{2}}-\delta_{\mu\nu}
\right)\partial_{\mu}(\rho u_{\nu})
\end{equation}
where the unit trace property of $S_{\alpha\beta}$ has been used.
In order to obtain an expression for $\Pi_{\alpha\beta}^{(1)}$ we
use equations (\ref{eq:collisionOperatorLBGKForm}) and (\ref{
eq:CE1DerivativeForm}) to give, to first order in the velocity,
\begin{equation}\label{eq:NonEquilibriumgDistribution}
g_{i\alpha\beta}^{(1)}= - \tau_{i}\left[t_{i}\left(
\frac{c_{i\mu}c_{i \nu}}{c_{s}^{2}}-\delta_{\mu\nu}
\right)\partial_{\mu}(\rho S_{\alpha\beta} u_{\nu})- \delta
m_{i\alpha\beta}- \delta p_{i\alpha\beta}\right]
\end{equation}
where we have made the approximation, correct to $O(\delta^{2})$,
that $g_{i\alpha\beta}^{(neq)} \simeq g_{i\alpha\beta}^{(1)} $.
The term $\delta m_{i\alpha}$ does not contribute to
$\Pi_{\alpha\beta}^{(1)}$, as can be seen from equation
(\ref{eq:PiDefinition}), since it is included in the rest mass.
It can further be shown that the contribution from the term
$\delta p_{i\alpha\beta} $ is essentially zero. We therefore find
\begin{eqnarray}\label{eq:PiExpansion}
 \lefteqn{ \Pi_{\alpha\beta}^{(1)}= -\tau_{0} \sum_{i}}\nonumber\\ & &t_{i}c_{i\alpha}c_{i\beta}(1+\eta_{0}c_{i\mu}c_{i\nu}\delta_{\mu\nu}+
\eta_{2}c_{i\mu}c_{i\nu}Q_{\mu\nu})  \left[\left(
\frac{c_{i\tau}c_{i
\varepsilon}}{c_{s}^{2}}-\delta_{\tau\varepsilon}
\right)\partial_{\tau}(\rho u_{\varepsilon})\right]
\end{eqnarray}
where we have substituted for the anisotropic relaxation
parameter, $\tau_{i}$. Assuming that the velocity tensors
\begin{equation}\label{eq:velocityTensor}
 E^{(n)}_{\alpha_{1}\ldots\alpha_{n}}=\sum_{i}t_{i}c_{i\alpha_{1}}\ldots\ldots c_{i\alpha_{n}}
\end{equation}
are isotropic up to sixth order, we recover the required form of
the tensor coupling between the order tensor and the velocity
gradient tensor.

Using these results it can be shown that,  apart from the term
associated with $\beta_{1}$, the required momentum evolution
equation (\ref{eq:qianMomentum}) is recovered when equation
(\ref{eq:FAlphaBeta}) is used as the momentum forcing term. The
term in $\beta_{1}$ could be recovered by including a fourth order
velocity product in the anisotropic relaxation parameter but this
would require the velocity tensors to be isotropic to eighth
order. However, for simplicity, the term in $\beta_{1} $ is
omitted in the results presented in this paper.

The order tensor evolution equation is recovered from the two
equations (\ref{eq:ChapmanEnskog1}) and (\ref{eq:ChapmanEnskog2})
by summing over the index $i$ and adding the two resulting
equations to give
\begin{equation}\label{eq:CEOrderEvolution}
 \partial_{t,0}(\rho Q_{\alpha\beta})+\partial_{t,1}(\rho
 Q_{\alpha\beta})+\partial_{\gamma}(\rho Q_{\alpha\beta} u_{\gamma})=\sum_{i}\chi_{i\alpha\beta}
\end{equation}
Equation (\ref{eq:CEOrderEvolution}) combined with the forcing
term (\ref{eq:angularForcing}) gives the required evolution
equation (\ref{eq:qianOrder}). However, in order to obtain the
result (\ref{eq:CEOrderEvolution}) it is necessary to suppress a
contribution of the form
\begin{equation}\label{eq:AngularCondition}
 \partial_{\gamma}(\sum_{i}c_{i\gamma}g_{i\alpha\beta}^{(1)})
 \end{equation}
at second order in the Chapman Enskog analysis and this is
achieved using the term associated with the factor $- 2$ in
equation (\ref{eq:LBGKcorrectionsMomentum}).

The term (\ref{eq:AngularCondition}) arises because conservation
of mass and momentum
 are alone insufficient to constrain all the first order elements of the distribution
function. This arises because in order to recover the required
macroscopic equations, the form of the equilibrium distribution
function must satisfy equations
(\ref{eq:EquilibriumDistributionFunctionFirstMoment}) .  However,
the limitations on the definitions of the moments of the
distribution function lead to the result
\begin{equation}\label{eq:FirstMomentConflict}
  \sum_{i}c_{i \gamma}g_{i\alpha\beta}\neq\rho S_{\alpha\beta}u_{\gamma}
\end{equation}
and as a consequence the argument of the derivative in
(\ref{eq:AngularCondition}) is non-zero.

\subsection{Choice of LB Parameters}
\label{subsec:ChoiceOfLBParameters}  The following correspondence
is found between the parameters of the LB algorithm and the
parameters of the target scheme
\begin{eqnarray}\label{eq:parameterMapping}
\beta_{4} & = & \frac{\mu_{2}^{2}}{4 \mu_{1}}-\rho
c_{s}^{2}\left(1- 2 \tau_{0}\zeta \right)\nonumber\\ \beta_{5}& =
& -\frac{\mu_{2}}{2}+\eta_{2} \rho \tau_{0}c_{s}^{2}\nonumber\\
\beta_{6}& = &\frac{\mu_{2}}{2}+\eta_{2} \rho \tau_{0}c_{s}^{2}
\end{eqnarray}
where the parameter $\zeta= 1 + \eta_{0}(1 + d/4 )$ with $d$ being
the dimensionality of the LB scheme which is taken to be 2 for the
results presented in Section (\ref{sec:Results}). The viscosity
set, $\{\beta_{4}, \beta_{5},\mu_{1},\mu_{2}\}$, of the Qian
scheme is recovered from the parameter set,
$\{\tau_{0},\eta_{0},\eta_{2}, \mu_{1},\mu_{2} \}$, of the LB
algorithm. We note from Qian(\cite{qian:98a}) that
$\beta_{6}-\beta_{5}=\mu_{2}$ and this is seen to be consistent
with the last two of equations (\ref{eq:parameterMapping}).  There
is one free parameter in the LB scheme since we require to recover
only four Qian parameters from five LB parameters. The free
parameter is taken to be $\eta_{0} $ which is adjusted to place
the LB scheme in a stable region of its parameter space, as is
explained below.

The equations (\ref{eq:parameterMapping}) can be inverted to give
\begin{eqnarray}\label{eq:invertedParameterMapping}
\eta_{2} &=& -\frac{4 \mu_{1} \zeta (2 \beta_{5}+\mu_{2})
}{\mu_{2}^{2}-4 \mu_{1}(\beta_{4}+\rho c_{S}^{2})}\nonumber \\
\tau_{0}& = & \frac{4 \beta_{4}+4 \rho
c_{s}^{2}-\frac{\mu_{2}^{2}}{\mu_{1}}}{8 \rho c_{s}^{2}\zeta}
\end{eqnarray}
In the absence of any anisotropic terms, the last expression in
equation (\ref{eq:invertedParameterMapping}) becomes
$\beta_{4}/\rho=c_{s}^{2}(1-2 \tau_{0})$, which is equivalent to
the standard result for the kinematic viscosity in an LBGK scheme.
The relationship between the Qian parameters $\{\beta_{4},
\beta_{5},\mu_{1},\mu_{2}\}$ and the standard Leslie Coefficients
is given in Section \ref{sec:AnalyticalResults}.

We now establish an approximate criterion for the stability of the
LB algorithm by considering a system which is evolving towards a
uniform density and velocity distribution at all points.  The
algorithm involves repeated application of the collision operator,
$M_{i\alpha\beta j \mu\nu} \equiv \underline{\underline{M}}$, and
ignoring forcing terms, the non-equilibrium part of the
distribution function is transformed at each time step according
to the symbolic equation
\begin{equation}\label{eq:symbolicBoltzmannEquation}
  \underline{g}^{(neq)}\rightarrow(\underline{\underline{1}}+
  \underline{\underline{M}})\cdot\underline{g}^{(neq)}
\end{equation}
In order for the non-equilibrium part to decay to zero under
successive applications of equation
(\ref{eq:symbolicBoltzmannEquation}),  it is necessary that
\begin{equation}\label{eq:stabilityCondition}
  -1 < (\underline{\underline{1}}+\underline{\underline{M}})_{i\alpha\beta j \mu\nu}< 1
\end{equation}
for all choices of indices.  This equation may be recast as the
stability criteria
\begin{equation}\label{eq:StabilityCriterion}
-2<M_{(Diag)}<0  ~~~~~~~~~~~~~~~~~~~ |M_{(OffDiag)}|<1
\end{equation}

However, it should be remembered that the matrix
$\underline{\underline{M}}$ is a function of position, as a
consequence of its dependence on the order tensor through the
anisotropic relaxation parameter, $\tau_{i}$.  Hence, in a given
simulation, the conditions (\ref{eq:StabilityCriterion}) cannot be
guaranteed under all flow conditions. The criteria have been
achieved in the implementation of the algorithm described in
Section \ref{sec:Results} by explicit evaluation of all the
non-zero elements of the collision matrix prior to running the
algorithm and  after making appropriate assumptions about the
ordering in the system. The parameter $\eta_{0}$ is then chosen to
ensure that the conditions (\ref{eq:StabilityCriterion}) are
satisfied for all possible values of the director and scalar order
parameter.
\section{Analytical results from the Qian formalism}
\label{sec:AnalyticalResults} In this section we quote without
detailed proof some analytical results which can be derived from
the Qian formalism and which will be used in Section
\ref{sec:Results} to validate the algorithm.  Further discussion
of the results and their significance is being prepared for a
future publication. The results quoted in Section
\ref{sec:Results} have been derived for a system in which the
director is confined to lie in a two dimensional plane and the
order tensor for such a system may be written in the form
\begin{equation}\label{eq:2DOrderTensor}
  Q_{\alpha\beta}=S (2 n_{\alpha} n_{\beta}-\delta_{\alpha\beta})
\end{equation}
where the time and position dependent quantities $S$ and
$n_{\alpha}$ are the order parameter and director. If the form
(\ref{eq:2DOrderTensor}) is substituted into the free energy
expression (\ref{eq:LandauFreeEnergy}) and all the time and
spatial derivatives are removed, we find the equilibrium
Landau-deGennes free energy  in the presence of a uniform magnetic
field is given by
\begin{equation}\label{eq:equilbriumFreeEnergy}
F_{LdG}^{(equ)} = \alpha_{F} S_{H}^{2}+4 \gamma
S_{H}^{4}-\frac{1}{2}S_{H}\chi_{a}H_{\alpha}^{2}
\end{equation}
where it is assumed that the director is parallel to the magnetic
field. The equilibrium order parameter, which minimises equation
(\ref{eq:equilbriumFreeEnergy}), has the form
\begin{equation}\label{eq:orderParameterInMagneticField}
S_{H}=S_{+}+S_{-}
\end{equation}
with
\begin{equation}\label{eq:SRoots}
S_{\pm}=\frac{1}{12}\left(\frac{27 \chi_{a} H^{2}}{\gamma}\pm 3
\sqrt{3}\sqrt{\frac{8 \alpha_{F}^{3}+27 \gamma\chi_{a}^{2}H^{4}
}{\gamma^{3}}}\right)^{1/3}
\end{equation}
where the free energy parameter, $\alpha$, will be negative in the
nematic phase and we choose the real solutions of $S_{0}$.  In the
limit the $H\rightarrow 0$, we have
\begin{equation}\label{eq:OrderParameterZeroH}
  S_{0}=\frac{1}{2}\sqrt{\frac{-\alpha_{F}}{2 \gamma}}
\end{equation}
which is the equilibrium order parameter which effectively
minimises the Landau-deGennes free energy in the absence of
director gradients.  Using equation (\ref{eq:2DOrderTensor}) with
$S$ set to $ S_{0}$ we may follow the arguments given in the
Appendix (B) of Qian to determine the relationship between the
Leslie coefficients of a material and the viscosity coefficients
used in the Qian scheme.  It is found for a system with the order
tensor given by (\ref{eq:2DOrderTensor})
\begin{equation}
\begin{array}{rclcrcl}\label{eq:betaCoefficients}
\beta_{1}  & = &\alpha_{1}/(4 S_{0}^{2})& &  \beta_{4} & =
&(1/2)\left(2
\alpha_{4}+\alpha_{5}+\alpha_{6} \right)\\

\beta_{5}& = & \alpha_{5}/(2 S_{0})& & \beta_{6}& = & \alpha_{6}/(2 S_{0})\\

\mu_{1}& = & (\alpha_{3}-\alpha_{2})/(8 S_{0}^{2})& &  \mu_{2}&=&
(\alpha_{2}+\alpha_{3})/(2S_{0})
\end{array}
\end{equation}
where $S_{0}$ is the order parameter defined by equation
(\ref{eq:OrderParameterZeroH}). In simple shear flow between two
parallel plates and in the absence of any external field, equation
(\ref{eq:qianOrder}) for the evolution of the order tensor can be
solved to find the steady state value of the angle of orientation
and the order parameter.  In the centre of the flow, remote from
the walls, the gradients in the order tensor may be ignored and
the director is found to lie at an angle $\theta$ with respect to
the flow velocity where
\begin{equation}\label{eq:directorAngle}
  \cos(2\theta)=-\frac{4 S \mu_{1}}{\mu_{2}}
\end{equation}
and this is directly equivalent to the result found for the ELP
equations, $\cos(2 \theta)=-\gamma_{1}/\gamma_{2}$ \cite{
deGennes:93a}.  The order parameter is also modified in shear flow
and is given by the solution of the cubic equation
\begin{equation}\label{eq:OrderParameterCubic}
a x^3 + b x^2 + c x + d  =  0
\end{equation}
where $ x = S^2$ and
\begin{equation}
\label{eq:OrderParameterShearFlow}
\begin{array}{lllclll}
 a & = & 1024 \gamma^2 &~~~~& b & = & 256 \alpha \gamma \\
 c & = & 16 (\alpha^{2}+(\mu_{1}\partial_{y}u_{x})^2) & ~~~~ &
 d & = & - (\mu_{2}\partial_{y}u_{x})^2 \\
\end{array}
\end{equation}
This equation may be solved and to first order in the velocity
gradient the order parameter is found to be
\begin{equation}\label{eq:deltaS0}
  S=S_{0} +\frac{\sqrt{\mu_{2}^{2}-16 S_{0}^{2}\mu_{1}^{2}}}{8
  |\alpha_{F}|}|\partial_{y}u_{x}|
\end{equation}
and hence  the order parameter increases as a consequence of flow
alignment. Finally we consider the Miesowicz viscosities in a
system with a variable order parameter.  The stress tensor
evolution equation can be solved for a simple shear flow in the
presence of a strongly aligning external field.  If the director
is given by $\{n_{x},n_{y}\}$, the associated Miesowicz viscosity,
$ \eta_{eff}$, is given by
\begin{eqnarray}\label{eq:MiesowiczViscosity}
  \eta_{eff} & = & \frac{1}{4
  S_{0}^{2}}\left\{S_{0}^{2}[2\alpha_{4}+\alpha_{5}+\alpha_{6}]
  +S^{2}[\alpha_{3}-\alpha_{2}+4 \alpha_{1} n_{x}^{2}n_{y}^{2}]\right.\nonumber \\ & &\left.
  + S S_{0}[\alpha_{6}(2 n_{x}^{2}-1)+(\alpha_{2}+\alpha_{3})(n_{x}^{2}-n_{y}^{2})
  + \alpha_{5}(2n_{y}^{2}-1)] \right\}
\end{eqnarray}
where $S$ is the order parameter in the flow and $S_{0}$ is the
equilibrium value of the order parameter defined in equation
(\ref{eq:OrderParameterZeroH}). In the limit that the order
parameter is fixed ({\em ie} $S = S_{0}$), equation
(\ref{eq:MiesowiczViscosity}) reduces to the standard results
\begin{eqnarray}\label{eq:standardMiesowiczResults}
\eta_{par}& = &
\frac{1}{2}(\alpha_{4}+\alpha_{3}+\alpha_{6})\nonumber\\
\eta_{perp}& = & \frac{1}{2}(\alpha_{4}-\alpha_{2}+\alpha_{5})
\end{eqnarray}
where $\eta_{par}$ and $\eta_{perp}$ correspond to the director
being aligned parallel, ${\bf n}=\{1,0\}$ and perpendicular, ${\bf
n}=\{0,1\}$, to the flow.
 In the presence of both a velocity gradient and a strong aligning field the change in  the order parameter is dominated by the aligning field and is hence given to a good
approximation by equations
(\ref{eq:orderParameterInMagneticField}) and (\ref{eq:SRoots}).

\section{Results}
\label{sec:Results}

A number of simulations were undertaken in order to validate the
method described in the previous section.  All the results were
obtained using a $D2Q13$ lattice.

In the absence of flow and in the presence of periodic boundary
conditions, the order parameter is found to be consistent  with
the value predicted by equation (\ref{eq:OrderParameterZeroH}) to
machine accuracy.  If an external magnetic field is applied
through the introduction of a molecular field term of the form
(\ref{eq:MagenticMolecularField}), it is found that the director
becomes fully aligned with the external field and that the order
parameter follows equation
(\ref{eq:orderParameterInMagneticField}) with an accuracy of
better than $0.1\%$. These two results confirm that the free
energy correctly controls the dynamical equations through the
associated molecular field.  In principal, therefore, the effects
of temperature could be introduced into the model through the
coefficients in the Landau-deGennes free energy.

The angle of alignment of the director in a shear flow in the
absence of an external magnetic field is given by equation
(\ref{eq:directorAngle}).  In order to test that the technique
recovered this result, simulations were run with the free energy
parameters, $(\alpha_{F}=-0.0512,\beta_{F}=0,\gamma_{F}=0.01)$,
chosen to give an equilibrium order parameter of $0.8$ and the
parameters, ($L_{1}=0.001,L_{2}=0$), selected to give nematic
elastic behaviour equivalent to the one constant approximation.
The LB parameters were chosen to be, $(\tau_{0}=1.1000,
\eta_{0}=0.3036, \eta_{2}=0.1565, \mu_{1}=0.3823,s \mu_{2}=-1.261,
\rho=1.8)$. If the order parameter $S_{0}$ is assumed to be 0.8,
these values recover the viscosity ratios,
$(\alpha_{2}/\alpha_{4}=-0.9556, \alpha_{3}/\alpha_{4}=-0.0144,
\alpha_{5}/\alpha_{4}=0.5565, \alpha_{6}/\alpha_{4}=-0.4135)$,
which are consistent with MBBA at 25$^o$ C \cite{doorn:75a}

\begin{figure}[p]
  \centering
\resizebox{1.0
\linewidth}{!}{\includegraphics{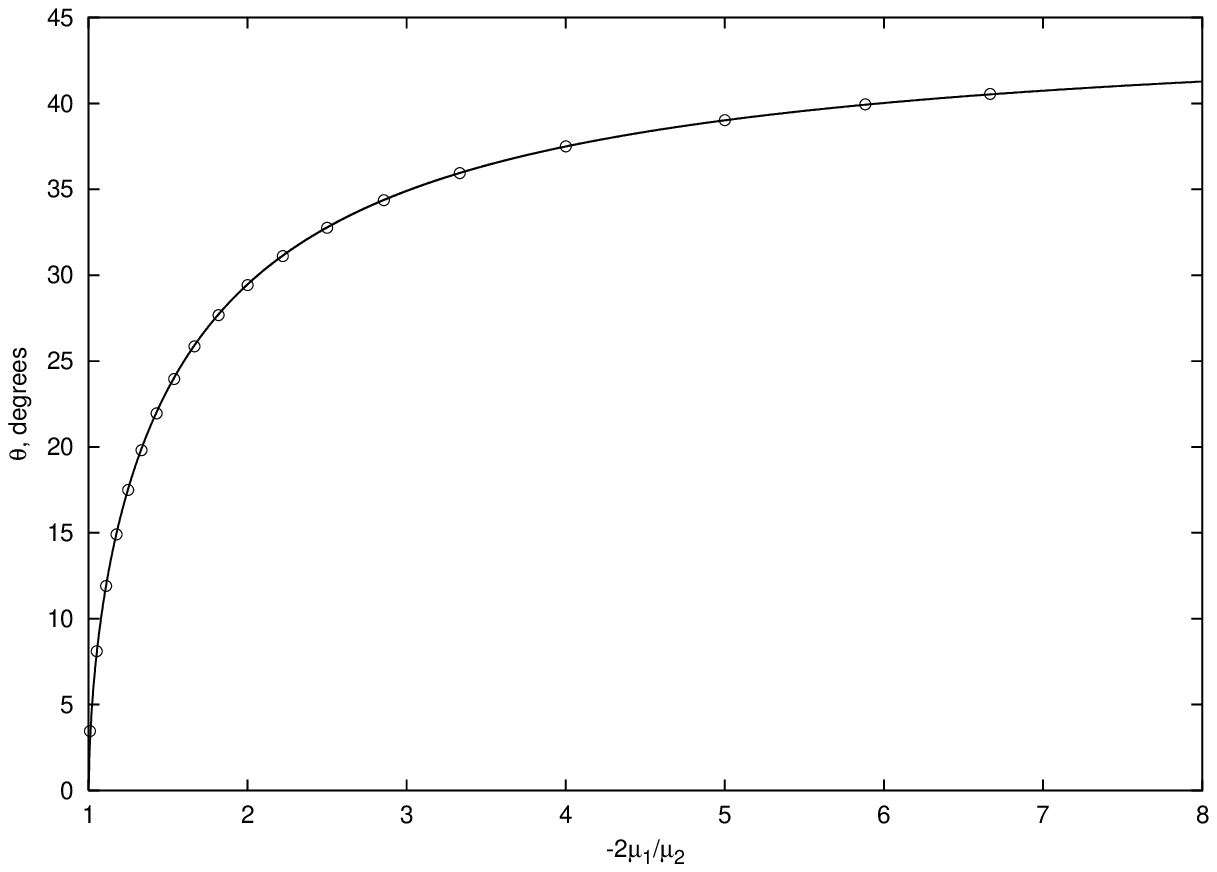}}
  \caption{Director orientation, $\theta$,  in a shear flow as
  a function of $\mu_{2}/\mu_{1}$.  The continuous curve
   is $\cos(2 \theta)=-4 S \mu_{1}/\mu_{2}$}
  \label{fig:FlowAlignement}
\end{figure}

With this choice of parameters, the angle of rotation of the
director is predicted to be $7^{o}$ and this is recovered to an
accuracy of less than $1\%$.  The parameters $\mu_{1}$ and
$\mu_{2}$ were then adjusted to give a range of values of angle of
alignment and the results are shown in Figure
(\ref{fig:FlowAlignement}) together with the expected curve,
equation (\ref{eq:directorAngle}), which is shown as a continuous
curve with the value of the order parameter taken from the
simulation data. The change in the order parameter predicted by
equation (\ref{eq:deltaS0}) is also recovered to within less than
$0.3\%$.

\begin{figure}[p]
  \centering
  \resizebox{1.0 \linewidth}{!}{\includegraphics{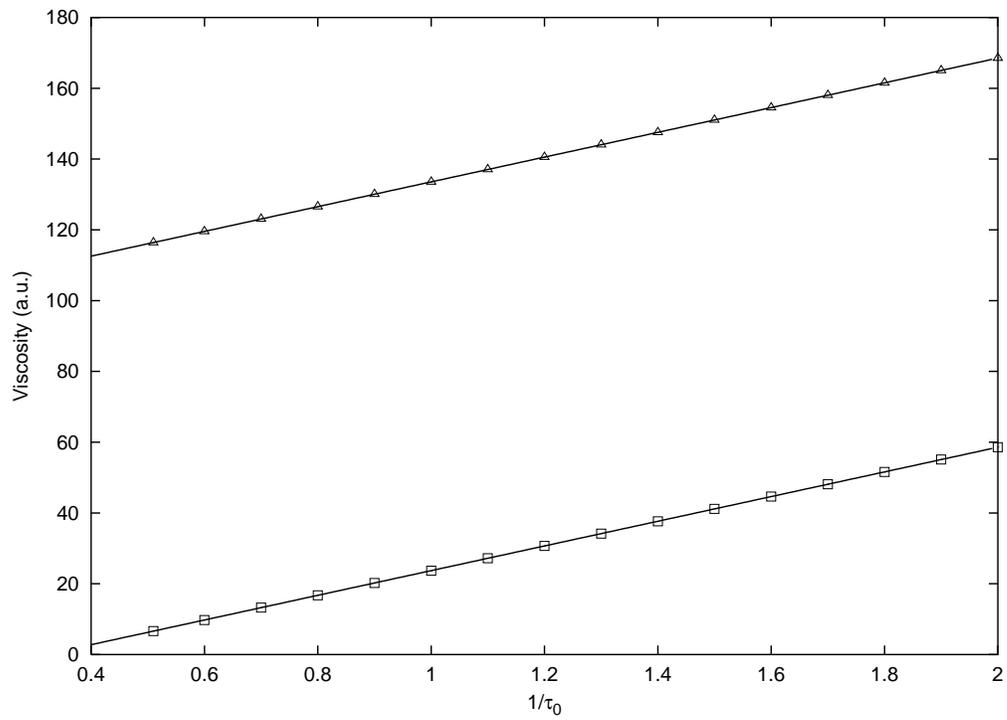}}
  \caption{Viscosity in shear flow as a function of $1/\tau_{0}$.  The upper (lower) line is for
  director perpendicular (parallel) to the flow direction.}
  \label{fig:MiesowiczViscosities}
\end{figure}

As a final test of the algorithm, the Miesowicz viscosities were
measured for simulations in which the director orientation was
controlled by a strong magnetic field.  Using the parameters
appropriate for MBBA the expected ratio of the viscosity with the
director aligned parallel and perpendicular to the direction of
flow is 5.03 and this value was recovered with an error of $
0.1\%$.  The expected form of the viscosity is given by equation
(\ref{eq:MiesowiczViscosity}) and it can be shown that this
implies that the viscosity should be a linear function of
$1/\tau_{0}$.  Figure (\ref{fig:MiesowiczViscosities}) shows that
the expected linearity is observed within the simulations;  the
slope to intercept ratio for each curve is in error by less than
$0.2 \%$

\section{Conclusions}
\label{sec:Conclusions}

In this paper we have presented a generalised lattice Boltzmann
scheme which recovers the tensor order parameter equations for a
nematic liquid crystal proposed by Qian and Sheng \cite{qian:98a}.
The generalised method is based on a single tensor density from
which all the macroscopic quantities can be recovered.  A Chapman
Enskog analysis demonstrates that the algorithm recovers the
target macroscopic equations and test simulations demonstrate that
the method correctly recovers the evolution of the director, the
order parameter and the velocity gradients in the presence of
shear and magnetic fields.

The method will generalise straightforwardly to three dimensions
and work is currently in progress to extend the scheme to model a
mixture of a isotropic and nematic fluids.

\begin{large}
\noindent {\bf Acknowledgements}
\end{large}
We thank Dr D J Cleaver for valuable conversations throughout the
developments described in this paper.

\newpage

\end{document}